\documentclass[doublespacing]{elsart}

\usepackage{graphicx}
\usepackage{epsfig}
\usepackage{amssymb}

\begin{document}
\begin{frontmatter}


\journal{SCES'2001: Version 1}


\title{Quantum Monte Carlo simulations of the t-Jz model with
stripes on the square lattice.}

%

\author{Jos\'e Riera\corauthref{1}} 

%
 
\address{CONICET and Universidad Nacional Rosario, Argentina}

%
%
%
%


%
%
%
%

\corauth[1]{jose@jaguar.fceia.unr.edu.ar}


\begin{abstract}

We have performed finite temperature quantum Monte Carlo
simulations on the t-Jz model on the square lattice. An
on-site potential, representing the effect of external
mechanisms, is used to stabilize a state of straight
site-centered stripes.
We show in the first place that various results of
our simulations can be related to features observed in
high-Tc superconductors giving validity to our model. In
particular, it is shown that only below a certain
temperature the spin regions between stripes are in
anti-phase corresponding to an ``incommensurate" magnetic
order.
Then, we examine the hole-hole correlations
concluding that in the presence of this kind of stripes
no sign of attraction of holes is seen at the
lowest temperatures we can reach. The consequences for
several theories regarding the relationship between stripes
and superconductivity are discussed.

\end{abstract}

%
%



\end{frontmatter}

%
%
%
%
%

The stripe order, experimentally seen in some underdoped 
cuprates,\cite{ichikawa} is one of the most intensely studied problems
in high-T$c$ superconductivity. In particular, the mechanism that
leads to the formation of stripes and the relation of stripes with
superconductivity are at the center of strong controversies.
Different mechanisms proposed imply different
charge and magnetic orderings as the temperature is lowered at a 
given doping.
In some theories, the mechanism of stripe formation is intrinsic
to the electronic interactions in two-dimensional (2D) Cu-O
planes.\cite{emery,white}
Alternatively, various groups have proposed the possibility of 
external effects such as electron-phonon couplings, lattice
anisotropy, Coulomb repulsion of out-of-plane ions, etc. which
could stabilize the stripe states leading to a picture closer
to experiments.\cite{eroles,prelovsek,castroneto,kampf}

These external mechanisms can be modeled by adding to the 2D $t-J$
model an on-site potential favoring the presence of holes on
the stripes. We have adopted this point of view and the resulting
model have been studied by finite temperature quantum Monte Carlo 
techniques (QMC) on an $8 \times 8$ cluster with 8 holes ($x=0.125$)
and on a $12 \times 12$ cluster with 12, 18 and 24 holes ($x=0.083$,
0.125 and 0.167), and
periodic boundary conditions.\cite{riera} We imposed
equally spaced site-centered stripes by taking the on-site
potential $-e_s$ ($e_s$) on (outside) the stripes. In order to
reduce the ``minus sign problem" of QMC simulations we have worked
essentially around the Ising limit of the exchange term
(anisotropy constant $\gamma=0$). We adopted 
$J=0.35$, $t=1$, parameters generally taken to describe the
physics of Cu-O planes and $J=0.7$ to exaggerate magnetic
effects. In particular pairing would be more favorable with
this larger value of $J$.

\begin{figure}
\begin{center}
\epsfig{file=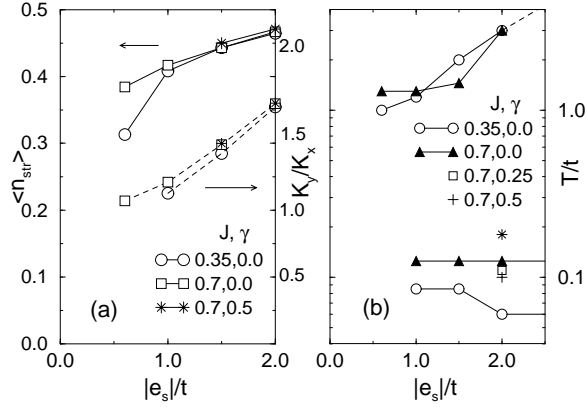,width=5.5cm,angle=-90}
\end{center}
\caption{
(a) Hole density on stripes and ratio of the kinetic energies
along and transversal to the stripes as a function
of the on-site potential.
(b) Phase diagram in the temperature-on-site potential plane for
the $8\times 8$ cluster. The curves at high (low) T correspond to
a crossover in the charge (magnetic) structure factor. The star
corresponds to $12 \times 12$, $x=0.083$, $J=0.7$, $\gamma=0.0$.}
\label{fig1}
\end{figure}

The effects of the on-site potential are shown in Fig.~\ref{fig1}(a).
A similar range of $e_s$ was considered 
previously.\cite{eroles,prelovsek}
Notice first that quite large values of $e_s$ are necessary to obtain 
$\langle n_{str} \rangle \approx 0.4$. However, even for 
these values
of $e_s$, we show that there is a considerable kinetic energy in the 
direction transversal to the stripes as seen experimentally.\cite{zhou}
These results correspond to the lowest reachable temperature. The
behavior of these quantities is very smooth with $T$.
In the phase diagram of Fig.~\ref{fig1}(b)
we indicate a high-temperature crossover in the
charge sector and a low-temperature crossover in the spin sector.
This sequence of orderings is in agreement with experimental
observations.\cite{ichikawa,hunt} These crossovers correspond to
changes in the weights of the peaks of charge and magnetic structure
factors. At a high $T$, the most weighted peak shifts from
$(\pi,0)$ to $(2\delta,0)$ ($\delta=n\pi/L$, n: number of stripes, 
$L=8$, 12) in the charge 
structure factor, and at low $T$ the most weighted peak
shifts from $(\pi,\pi)$ to $(\pi \pm \delta,\pi)$ in
the spin sector.

\begin{figure}
\begin{center}
\epsfig{file=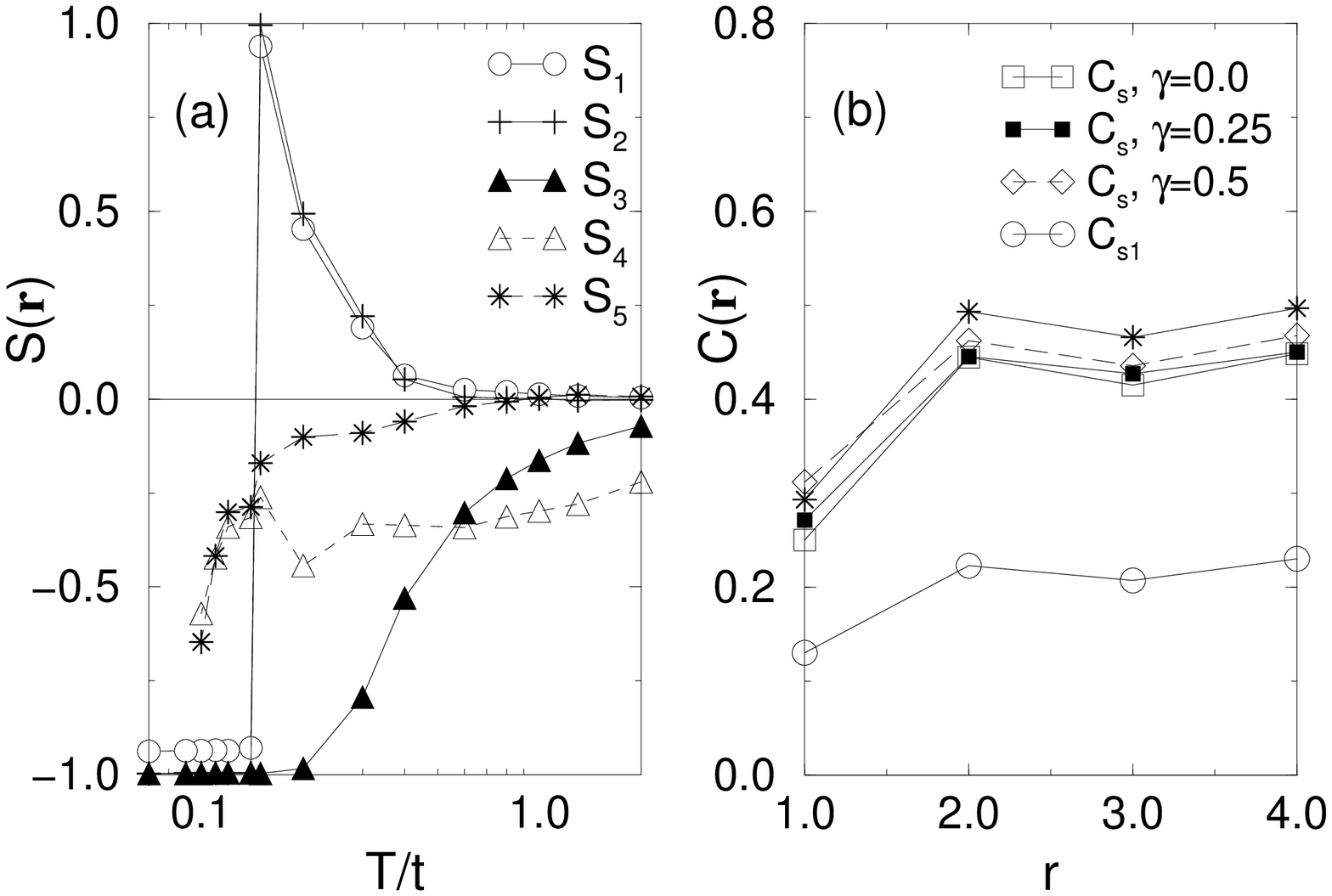,width=5.5cm,angle=-90}
\end{center}
\caption{
(a) Spin-spin correlations (defined in the text) vs. temperature
on $8\times 8$ cluster, and
$J/t=0.7$, $e_s=1.5$, $\gamma =0$. S$_5$ and S$_6$ have been
multiplied by 5.}
(b) Hole-hole correlations as a function of distance.
C$_s$: along the stripe; C$_{s1}$: between a site
  on the stripe and a site on the column next to it.
C$_{s1}$ has been multiplied by 10.
  The stars indicate exact results for an eight site
  $t-Jz$ ring at quarter filling.
\label{fig2}
\end{figure}

A more detailed description of the low-$T$ crossover is given
in Fig.~\ref{fig2}(a). In the $8 \times 8$ cluster, the spin-spin 
correlations on the center leg of the spin 3-leg ladder (S$_3$, 
$r=|{\bf r}_i-{\bf r}_j|=1$) have a rapid growth as the temperature
decreases and once they reach their maximum value these ladders
become in anti-phase as $T$ is further reduced. This is apparent
in the correlations across the stripe S$_1$ and S$_2$ ($r=2$ and
4). This anti-phase domain ordering leads to the incommensurate
peak mentioned before. Notice that the hole density on
the stripe has already achieved its maximum value at a temperature
much higher than the one at which this magnetic ordering occurs.
The correlations along the stripe (S$_4$ and
S$_5$, $r=1$ and 2) are much smaller and they do not show a
strong dependence on $T$. The behavior shown in this figure is
typical of all the parameters and clusters examined.

Finally, in Fig.~\ref{fig2}(b) we show the hole-hole correlations
$C({\bf r})$ for the same parameters and measured at the
lowest reachable temperature (typically $\approx 0.1 t$). These
correlations are smallest
at the shortest distance indicating lack of attraction and their 
overall behavior is reminiscent of a metallic state. 
We also show $C({\bf r})$ for the $t-Jz$ model on
an eight-site ring at quarter filling. It can be seen that
these correlations are virtually identical to the previous ones
except for a trivial shift due to the fact that the hole density
on the stripes is slightly less than quarter filling. Similar
behavior is observed when an XY spin term (up to $\gamma=0.5$) is 
included in the Hamiltonian. Besides, exact diagonalization 
results on $t-J$ chains show very little dependence on $\gamma$
at quarter filling.

The main features shown in Figs.~\ref{fig1}(b) and ~\ref{fig2}(b)
are consistent with a number of experiments and they indicate that
the spin ordering is not the driving force for the formation of
stripes but are a {\em consequence} of charge inhomogeneity.
This argues in favor of an external mechanism favoring stripes.
The results for the hole-hole correlations indicate lack of hole
pairing on the stripes. Hence, close to the Ising limit, 
site-centered stripes 
would compete with superconductivity and a coexistence between a 
stripe and superconductivity would correspond to 
macroscopically separated stripe and superconducting 
phases. However, we do not exclude the possibility of pairing in
{\em bond-centered} stripes.

%

%
%
%



\begin{thebibliography}{00}

\bibitem{ichikawa} N. Ichikawa, {\it et al.}, Phys. Rev. Lett. 
        {\bf 85}, 1738 (2000).

\bibitem{emery} V. J. Emery, {\it et al.}, Phys.~Rev.~B {\bf 56},
        6120 (1997).

\bibitem{white} S. R. White and D. J. Scalapino, Phys. Rev. B
     {\bf 61}, 6320 (2000).

\bibitem{eroles} J. Eroles, {\it et al.},
      Europhys. Lett. {\bf 50}, 540 (2000).

\bibitem{prelovsek} P. Prelovsek, {\it et al.}, Phys. Rev. B
       {\bf 64}, 052512 (2001).

\bibitem{castroneto} A. H. Castro Neto, cond-mat/0102281.

\bibitem{kampf} A. P. Kampf, {\it et al.}, cond-mat/0102554.

\bibitem{riera} J. A. Riera, Phys.~Rev.~B {\bf 64}, 104520
       (2001).

\bibitem{zhou} X. J. Zhou, {\it et al.}, Science {\bf 286}, 268
               (1999).

\bibitem{hunt} A. W. Hunt, {\it et al.}, Phys. Rev. Lett. {\bf 82},
               4300 (1999).

\end{thebibliography}
\end{document}